\documentclass[twocolumn,showpacs,preprintnumbers,amsmath,amssymb]{revtex4}

\usepackage{epsfig}
\usepackage{graphics}
\usepackage{graphicx}
\usepackage{dcolumn}
\usepackage{bm}

\begin{document}

\preprint{APS/123-QED}

\title{Inverse synchronizations in coupled time-delay systems with inhibitory coupling}

\author{D.~V.~Senthilkumar$^{1,2}$}
\email{skumar@cnld.bdu.ac.in}
\author{J. Kurths$^{2,3}$}%
 \email{Juergen.Kurths@pik-potsdam.de}
\author{M.~Lakshmanan$^4$}%
 \email{lakshman@cnld.bdu.ac.in}
\affiliation{%
$^1$Centre for Dynamics of Complex Systems, 14469 Potsdam Germany\\
}%
\affiliation{%
$^2$Potsdam  Institute for Climate Impact Research, 14473 Potsdam Germany\\
}%
\affiliation{%
$^3$Humboldt University, Berlin, Germany\\
}%
\affiliation{%
$^4$Centre for Nonlinear Dynamics,Department of Physics,
Bharathidasan University, Tiruchirapalli - 620 024, India\\
}%

\date{\today}

\begin{abstract}

Transitions between inverse anticipatory, inverse complete and inverse lag
synchronizations are shown to occur as a function of the coupling delay in unidirectionally
coupled time-delay systems with inhibitory coupling. We have also
shown that the same general asymptotic stability condition obtained using the Krasovskii-Lyapunov
functional theory can be valid  for the cases where (i) both the coefficients of
the $\Delta(t)$ (error variable) and $\Delta_\tau=\Delta(t-\tau)$ (error variable with delay) terms 
in the error equation corresponding to the
synchronization manifold are time independent and (ii) the coefficient of the
$\Delta$ term is time independent while that of the $\Delta_\tau$ term is time
dependent. The existence of different kinds of synchronization are 
corroborated using similarity function, probability of synchronization and also
from changes in the spectrum of Lyapunov exponents of the coupled
time-delay systems. 

\end{abstract}

\maketitle

{\bfseries Synchronization of chaotic oscillations is one of the most
fundamental phenomena exhibited by coupled chaotic oscillators. Since the
identification of chaotic synchronization in identical systems,
different kinds of synchronizations such as generalized, phase, lag,
anticipatory and intermittent synchronizations, and their variants, have been 
reported in the literature. Recent studies on synchronization have also
been focused on the existence of inverse synchronization. In this connection,
we have identified different kinds of inverse synchronization and  transitions
among them as a function of
a single parameter in coupled time-delay systems with inhibitory coupling.
It is now recognized that inhibitory couplings are important and widespread in nature,
specifically in biological systems, neurophysiology and in many natural networks.
We have also obtained a suitable stability condition for asymptotic stability of
the synchronized state using Krasovskii-Lyapunov functional theory. The results are 
corroborated using similarity function, probability of synchronization and from the
changes in the spectrum of Lyapunov exponents of specific coupled time-delay systems.}

\section{\label{sec:level1}Introduction}

Synchronization is an interesting dynamical phenomenon exhibited by interacting 
oscillators in diverse areas of science and technology~\cite{asp2001,sbjk2002}.
It has become an area of active research since the identification of
synchronization in chaotic oscillators~\cite{hfty1983}. In recent years,
different types of synchronization and generalizations have been reported
both experimentally  and theoretically~\cite{asp2001,sbjk2002,hfty1983,skhck1995,lkup1995,
rblk2000,sblmp2001,nfr1995,lkup1996,rb1998,mgrasp1996,
ytycl1997,mgrasp1997,srik2002,mzgww2002,huv2002,huv2001,cm2001,zyk2005,ssip2001,
iwup2001,iwup2002,aukh2003,emsss2002,
emsran2006,emsran2004,dwsag2006,hzbc2007,jmxyw2007}.  Among them,
inverse or anti-synchronization~\cite{ssip2001,iwup2001,iwup2002,aukh2003,emsss2002,
emsran2006,emsran2004,dwsag2006,hzbc2007,jmxyw2007} constitutes an important class of
synchronization, which is a typical feature of dynamical systems interacting
through inhibitory coupling.

By definition, inverse synchronization is the phenomenon~\cite{ssip2001,iwup2001,iwup2002,
aukh2003,emsss2002,emsran2006,emsran2004,dwsag2006,hzbc2007,jmxyw2007} where the state vectors, 
say the drive $x(t)$ and the response $y(t)$, of the synchronized systems have the same 
absolute values but of opposite signs, that is $x(t) = -y(t)$. While this is inverse complete synchronization, one may also identify inverse 
anticipatory synchronization where the response anticipates the drive, that is $y(t) = -x(t+\tau)$, 
where $\tau > 0$ is a constant. Similarly, when the response system lags the drive, one has inverse 
lag synchronization, that $y(t) = -x(t-\tau)$, $\tau > 0$. Just as in the case of regular synchronization, 
especially in nonlinear dynamical systems with inhibitory coupling, one may encounter all the above
types of inverse synchronization for appropriate ranges of the parameters.

The importance of inhibitory  or repulsive coupling is well acknowledged in
biological systems.  It is a  well established fact that couplings between neurons
are both excitatory and inhibitory~\cite{avrdc2006}. Ecological webs typically
have both positive and negative connections between their
components~\cite{xcjec2001}. Coupled lasers with negative couplings have also
been widely studied~\cite{twcmlt2006}.  The well known Swift-Hohenberg and
Kuramoto-Sivashinsky equations include such a term~\cite{mccpch1993}.  Currently,
it has also been realized that a large class of natural networks also have
inhibitory interactions among the nodes~\cite{sbvl2006,aaadg2008}. 

The first experimental observation of inverse synchronization was demonstrated in
coupled semiconductor laser diodes~\cite{ssip2001}, in which it was established
that inverse synchronization was caused by nonresonant coupling between the drive
and the response lasers. It was also shown that switching between synchronization and 
inverse synchronization was possible by slightly changing the pump current of
the drive laser~\cite{iwup2001}. Experimental observations and numerical simulations of 
synchronization and inverse synchronization of low frequency power drop-outs and 
jump-ups of chaotic semiconductor lasers were carried out in~\cite{iwup2002}. Inverse synchronization
was also observed both experimentally and numerically in unidirectionally coupled
laser systems with optical feedback~\cite{aukh2003,dwsag2006}, as well as
 in a class of chaotic delayed neural networks~\cite{hzbc2007} and
in coupled Ikeda systems with multi-feedback and multiple
time-delays~\cite{emsran2006}. Inverse anticipating synchronization 
was demonstrated in coupled Ikeda systems~\cite{emsss2002}. Further inverse
retarted/lag synchronization 
and the role of parameter mismatch were discussed in~\cite{emsran2004}. However it
may be noted that in none of the above studies the role of inhibitory coupling 
was investigated.

Despite the fact that a vast amount of literature is now available  on the phenomenon of
synchronization, inverse synchronizations have not been studied adequately, in
particular with inhibitory couplings between interacting dynamical systems.
In this paper, we report inverse synchronizations (inverse anticipatory,
inverse complete, and inverse lag synchronizations) in unidirectionally coupled
time-delay systems with inhibitory coupling.  We also present a sufficient
stability condition for asymptotic stability of the synchronized state following the Krasovskii-Lyapunov
functional approach for the cases where (i) the coefficients of the $\Delta(t)$ (error variable) and
$\Delta_\tau=\Delta(t-\tau)$ (error variable with delay) terms of the error equation corresponding to the synchronization
manifold are constant and (ii) the coefficient of the $\Delta$ term is constant and that
of the $\Delta_\tau$ term is time dependent. We show
that there is a transition from inverse anticipatory to inverse lag
synchronization through complete inverse synchronization as a function of the
delay time in the coupling. The tools to get these results are similarity
function, probability of synchronization and largest Lyapunov exponents of the
coupled time-delay systems.

The plan of the paper is as follows. In Sec.~II, we deduce a
sufficient condition for the asymptotic stability of the synchronized state for a system
of unidirectionally coupled scalar delay differential equations with inhibitory
coupling. We consider a piecewise linear delay differential equation as
an example for the case where the coefficients of both the $\Delta$ and
$\Delta_\tau$  terms in the error equation are constant and demonstrate the
existence of different types of inverse synchronization as a function of the
coupling delay in Sec.~III. Using the paradigmatic Ikeda system in Sec.~IV for the case where the
coefficient of the $\Delta_\tau$  term is time dependent while that of the
other is time independent, we show that the same general
stability condition is valid for the asymptotic stability of different types of inverse
synchronizations again as a function of the coupling delay. We also demonstrate these
dynamical transitions through numerical analysis. Finally, in
Sec.~V, we summarize our results.

\section{Coupled system and the stability condition}

Consider the following unidirectionally coupled drive, $x(t)$, and response,
$y(t)$, systems with inhibitory coupling of the form
\begin{subequations}
\begin{eqnarray}
\dot{x}(t)&=&-ax(t)+b_{1}f(x(t-\tau_{1})),  \\
\dot{y}(t)&=&-ay(t)+b_{2}f(y(t-\tau_{1}))-b_{3}f(x(t-\tau_{2})),
\end{eqnarray}
\label{eq.one}
\end{subequations}
where $b_1, b_2$ and $b_3$ are positive parameters, $a>0$, 
$\tau_1$ and $\tau_2$ are the feedback and the coupling delays,
respectively. The nonlinear function $f(x)$ is chosen to be a piecewise linear 
function which has been studied
in detail recently~\cite{dvskmlijbc,dvskml2005,dvskml2006},
\begin{eqnarray}
f(x)=
\left\{
\begin{array}{cc}
0,&  x \leq -4/3  \\
            -1.5x-2,&  -4/3 < x \leq -0.8 \\
            x,&    -0.8 < x \leq 0.8 \\              
            -1.5x+2,&   0.8 < x \leq 4/3 \\
            0,&  x > 4/3, \\ 
         \end{array} \right.
\label{eqoneb}
\end{eqnarray}

as the first example,  and 
\begin{equation}
f(x)=\sin(x_\tau)\equiv\sin(x(t-\tau)),
\label{eqonec}
\end{equation}
which is the well known Ikeda system~\cite{kihd1980}, as the second example.

Now the stability condition for the synchronization of the coupled time-delay
systems, Eqs.~(\ref{eq.one}), with the inhibitory delay coupling,
$-b_{3}f(x(t-\tau_{2}))$, can be obtained as follows. 
The time evolution of the
difference system (error function), associated with inverse synchronization, 
with the state variable $\Delta(t)=x_{\tau_2-\tau_1}+y(t)$, where $x_{\tau_2-\tau_1}=x(t-(\tau_2-\tau_1))$,
can be  written for small values of $\Delta$, by using the evolution Eqs.~(\ref{eq.one}), as
\begin{eqnarray}
\dot{\Delta}&=& \dot{x}_{\tau_2-\tau_1}+\dot{y}(t)\\
&=&-a\Delta+(b_1-b_2-b_3)f(x(t-\tau_{2}))+b_{2}f^\prime(x(t-\tau_{2}))\Delta_{\tau_1}.
\label{eq.difsys1}
\end{eqnarray}
The above evolution equation (\ref{eq.difsys1}) corresponding to the error function of the 
inverse synchronization manifold is inhomogeneous and so it is difficult to analyse
the system analytically. Nevertheless, the evolution equation can be written as a
homogeneous equation 
\begin{align}
\dot{\Delta}=-a\Delta+b_{2}f^\prime(x(t-\tau_{2}))\Delta_{\tau_1},
\label{eq.difsys}
\end{align}
for the specific choice of the parameters
\begin{align}
b_1=b_2+b_3. 
\label{para.con} 
\end{align}
Therefore, we will concentrate on this parametric choice.
The inverse synchronization manifold $\Delta=x_{\tau_2-\tau_1}+y=0$ corresponds to the following
distinct cases:
\begin{enumerate}
\item Inverse anticipatory synchronization occurs when $\tau_2 < \tau_1$ with $y(t)=-x(t-\hat{\tau}); \hat{\tau}=\tau_2-\tau_1<0$, where the state of the response system anticipates the inverse state of the drive system in a synchronized manner with the anticipating time $\hat{\tau}$ (whereas in the case of direct anticipatory synchronization, the state of the response system anticipates exactly the state of the drive
system, that is, $y(t)=x(t-\hat{\tau})$). 
\item Inverse complete synchronization results when $\tau_2 = \tau_1$ with $y(t)=-x(t); \hat{\tau}=\tau_2-\tau_1=0$, where the state of the response system evolves in a synchronized manner with
the inverse state of the drive system (whereas in the case of complete synchronization, the state of the response system evolves exactly identical to the state of the drive
system, that is, $y(t)=x(t)$). 
\item Inverse lag synchronization occurs when $\tau_2 > \tau_1$ with $y(t)=-x(t-\hat{\tau}); \hat{\tau}=\tau_2-\tau_1>0$, where the state of the response system lags the inverse state of the drive system in a synchronized manner with the lag time $\hat{\tau}$ (whereas in the case of direct lag synchronization, the state of the response system lags exactly the state of the drive
system, that is, $y(t)=x(t-\hat{\tau})$). 
\end{enumerate}

The synchronization manifold corresponding to Eq.~(\ref{eq.difsys}) is locally attracting 
if the origin of the above error equation is stable.
Following the Krasovskii-Lyapunov functional approach~\cite{nnk1963,dvskml2005}, we define a
positive definite Lyapunov functional of the form
\begin{align}
V(t)=\frac{1}{2}\Delta^2+\mu\int_{-\tau_1}^0\Delta^2(t+\theta)d\theta,
\label{klf}
\end{align}
where $\mu$  is an arbitrary positive parameter, $\mu>0$.  

It is to be noted that for the above error equation (\ref{eq.difsys}), the
coefficient of the $\Delta$ term is always constant while that of the  $\Delta_\tau$
term can be time dependent. Hence one can obtain two cases  depending on the choice
of the nonlinear functional form $f(x)$.  If the derivative of the function
turns out to be a constant as in the case of the piecewise linear function,
Eq.~(\ref{eqoneb}), one obtains a constant coefficient for the $\Delta_\tau$
term. On the other hand, if the derivative still depends on time as in
the case of the Ikeda
system, Eq.~(\ref{eqonec}), then the $\Delta_\tau$ term always has a time dependent coefficient. In the
following, we will show that the same  general stability condition derived from
the Krasovskii-Lyapunov approach can be valid for both the cases for the asymptotic 
stability of the different types of inverse synchronizations. However, there
also arises an even more general situation where the coefficients of both the $\Delta$
and $\Delta_\tau$ terms are time dependent. In this case the arbitrary positive 
parameter $\mu$ in the Lyapunov functional
is no longer a positive constant.  We have also designed a suitable
coupling form for this rather general situation, for which we can show that the
same general stability condition arrived from the Krasovskii-Lyapunov approach is 
still valid for the asymptotic stability of the synchronized state;
these results will be published elsewhere.

Note that from Eq.~(\ref{klf}), $V(t)$
approaches zero as $\Delta \rightarrow 0$.  Hence, the required solution $\Delta=0$ to the
error equation, Eq.~(\ref{eq.difsys}), is stable only when the
derivative of the Lyapunov functional $V(t)$  along the trajectory of
Eq.~(\ref{eq.difsys}) is negative. This requirement results in the condition for
stability as 
\begin{align}
\Gamma(\mu)=4\mu(a-\mu)>b_2^2f^{\prime}(x_1(t-\tau_2))^2.
\label{eq.ineq}
\end{align}
Again
$\Gamma(\mu)$ as a function of $\mu$ for a given $f^{\prime}(x)$ has an absolute
minimum at $\mu=a/2$ with  $\Gamma_{min}=a^2$.  Since $\Gamma\ge\Gamma_{min}= a^2$,
from the inequality (\ref{eq.ineq}), it turns out that a sufficient condition
for asymptotic stability is
\begin{align}
a>|b_2f^{\prime}(x(t-\tau_2))|.
\label{eq.asystab}  
\end{align}
This  general stability condition indeed corresponds to the stability condition
for inverse anticipatory, complete inverse and inverse lag synchronizations for
suitable values of the coupling delay $\tau_2$  corresponding to a fixed value
of the feedback delay $\tau_1$ for both the piecewise linear and Ikeda time-delay
systems corresponding to the cases where the coefficient of the $\Delta_\tau$
term in the error equation is time independent and time dependent, respectively. 

Further, it is interesting to note that if one substitutes $y\rightarrow \hat{y}=-y$
in Eq.~(\ref{eq.one}b), then the coupling becomes excitatory for the choice
of functional forms we have chosen. This is exactly the case we have studied in
~\cite{dvskml2005}, where direct anticipatory, complete and lag synchronizations exist
as a function of the coupling delay. However, one cannot obtain inverse (anticipatory,
complete and lag) synchronization
with excitatory coupling or direct (anticipatory, complete and lag) synchronization with
inhibitory coupling for the chosen form of the unidirectional nonlinear coupling
because of the nature of the parametric relation between $b_1, b_2$ and $b_3$ and the
stability condition (\ref{eq.asystab}).

Now from the form of the function $f(x)$ in
Eq.~(\ref{eqoneb}) for the piecewise linear time-delay system, one can obtain 
a less stringent stability condition as~\cite{dvskml2005}
\begin{align}
a>b_2,
\label{piecewise_stab}  
\end{align}
while 
\begin{align}
a>1.5b_2,
\label{piecewise_stab1}  
\end{align}
is the most general condition specified by (\ref{eq.asystab}) for
asymptotic stability of the synchronized state $\Delta=0$.

Correspondingly, one can obtain the stability condition for the coupled Ikeda systems as
\begin{align}
a>\vert b_2\cos(x(t-\tau_2))\vert.
\label{ikeda_stab}  
\end{align} 

\begin{figure}
\centering
\includegraphics[width=1.0\columnwidth]{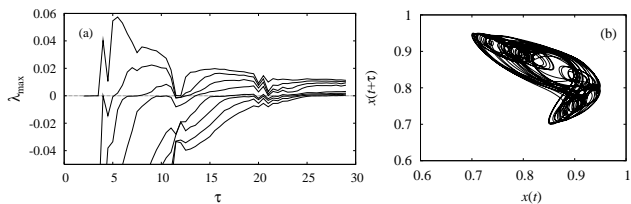}
\caption{\label{fig1} (a) The first ten maximal Lyapunov exponents
$\lambda_{max}$ of the piecewise linear time-delay system ((\ref{eq.one}a) and (\ref{eqoneb})) for the
parameter values $a=1.0, b=1.2,$  $\tau\in(2,29)$ and (b) Hyperchaotic
attractor for the  delay time $\tau=8.0$  with two positive Lyapunov
exponents for the above values of the other parameters. Here $\tau_1=\tau$.}
\end{figure}

\section{Inverse synchronizations in the coupled piecewise linear time-delay systems ((\ref{eq.one}) and (\ref{eqoneb}))}

Here, we will show the existence of inverse anticipatory, inverse complete and 
inverse lag synchronizations as a function of the coupling delay in the 
coupled piecewise linear time-delay systems (\ref{eq.one})-(\ref{eqoneb})
 with inhibitory coupling as an illustration for the case where the coefficients of both the
$\Delta$ and $\Delta_\tau$ terms in the error equation (\ref{eq.difsys}) are constant.

\begin{figure}
\centering
\includegraphics[width=1.0\columnwidth]{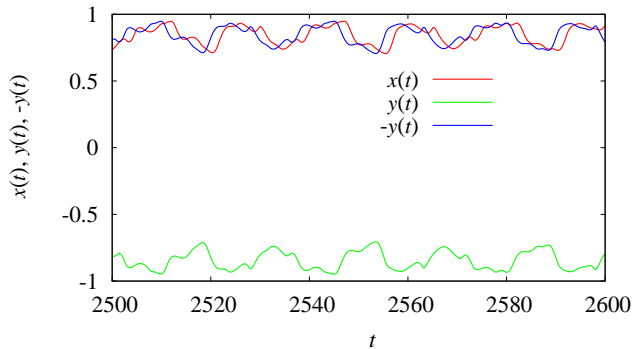}
\caption{\label{fig2} (Color online) The time trajectory  of the variables
$x(t), y(t)$ and $-y(t)$ for $a=1.0, b_1=1.2,b_2=0.6, b_3=0.6, \tau_1=8.0$  and $\tau_2=6.0$
of the coupled piecewise linear time-delay systems indicating IAS.}
\end{figure}

\subsection{\label{ias}Inverse anticipatory synchronization, IAS}

In this section, we present the existence of inverse anticipatory synchronization (IAS)
for the values of the coupling delay $\tau_2<\tau_1$, and for  fixed values of the other
system parameters. In particular, we have fixed  the parameters as
$a=1.0, b_1=1.2, \tau_1=8.0, \tau_2=6.0$ and the other two parameters $b_2$ and $b_3$ are fixed
according to the parametric condition~(\ref{para.con}). The first ten maximal Lyapunov
exponents $\lambda_{max}$ of the uncoupled piecewise linear time-delay system  in the range
of delay time $\tau\in(2,29)$ for the above choice of parameters are shown in Fig.~\ref{fig1}a.
It is clear from Fig.~\ref{fig1}a that for $\tau > 0.5$ at least two of the Lyapunov exponents are
positive and that the system is hyperchoatic. As an illustration, 
the hyperchaotic attractor with two positive Lyapunov exponents
for the value of the delay time $\tau_1=\tau=8.0$ is plotted in Fig.~\ref{fig1}b. 
The time trajectories of the variables $x(t), y(t)$ and $-y(t)$ of 
the coupled piecewise linear time-delay systems (\ref{eq.one})-(\ref{eqoneb})
are plotted  in Fig.~\ref{fig2} depicting the existence of IAS for $b_2=0.6$, for which the more
general stability condition, $a>1.5b_2$, is satisfied. It is evident from the figure that the
state of the response system $y(t)$ anticipates the inverse state of the drive system $x(t)$.
This is made visually clear by plotting the inverse state of the response, that is $-y(t)$. 
From Fig.~\ref{fig2}, it is clear that
the inverse state of the response ($-y(t)$)  anticipates the state of the drive $x(t)$, thereby 
illustrating the existence of IAS between the drive and the response systems.

\begin{figure}
\centering
\includegraphics[width=1.0\columnwidth]{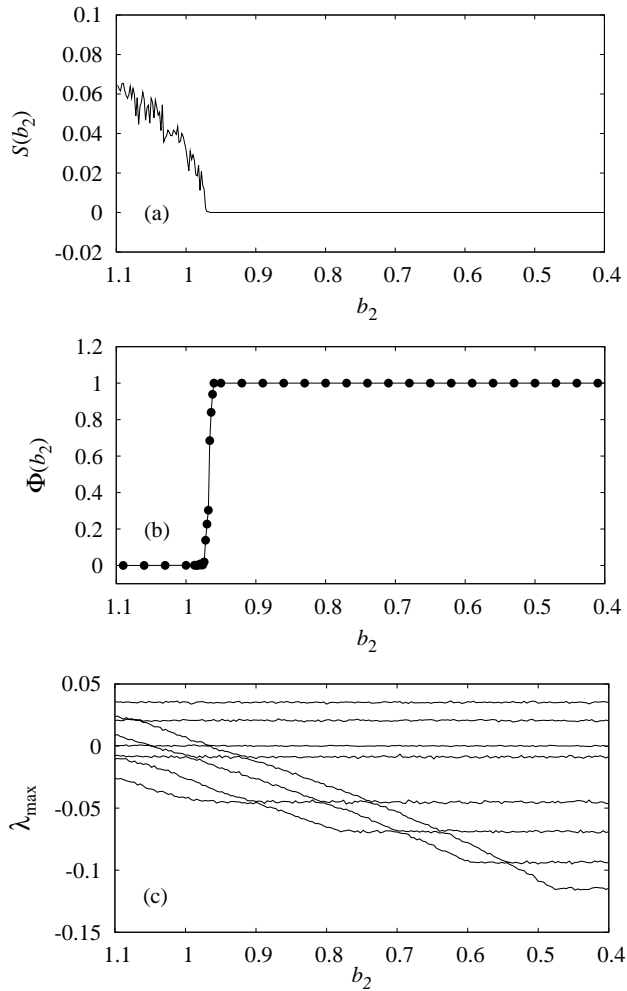}
\caption{\label{fig3} (a) Similarity function, $S_a(b_2)$, (b) Probability of
synchronization, $\Phi(b_2)$ and (c) First eight maximal Lyapunov exponents, $\lambda_{max}$,
of the coupled piecewise linear time-delay systems as a function of the
control parameter $b_2$, indicating the existence of IAS for $b_2<0.97$.}
\end{figure}

The existence of IAS for the above choice of
parameters is further confirmed using the notion of similarity function $S(b_2)$,
probability of synchronization $\Phi(b_2)$ and also from the spectrum of Lyapunov exponents of
the coupled time-delay systems. Now we use the notion of similarity function, 
introduced in~\cite{mgrasp1996} for characterizing 
the lag synchronization, to characterize the existence of IAS.
The similarity function, $S_a(b_2)$, is defined
as a time averaged difference between the variables $x(t)$ and $-y(t)$ (with mean
values being  subtracted) taken with the time shift $\tau$,
\begin{eqnarray}
S_a^2(b_2)=\frac{\langle[y(t-\tau)+x(t)]^2\rangle}
{[\langle x^2(t)\rangle\langle y^2(t)\rangle]^{1/2}},
\label{anti:sim}
\end{eqnarray}
where $\langle x \rangle$ means time average over the variable $x$. If
the signals $x(t)$ and $-y(t)$ are independent, the difference between them
is of the same order as the signals themselves.  If $x(t)$ = $y(t)$, as 
in the case of complete synchronization, the similarity function reaches a 
minimum, $S_a(b_2) = 0$, for $\tau = 0$.  On the other hand, if $S_a(b_2) = 0$ for the case 
$\tau\ne 0$, there exists a time shift $\tau$
between the two signals $x(t)$ and $-y(t)$ such that $y(t-\tau) = -x(t)$,
demonstrating inverse anticipatory synchronization. The similarity function,
$S_a(b_2)$, as a function of the parameter $b_2$ is shown in Fig.~\ref{fig3}a, 
which clearly indicates that the value of $S_a(b_2)$ oscillates with finite
amplitude ($S_a(b_2)>0$) above the value of the control parameter $b_2\approx 0.97$, indicating
the desynchronized evolution between the response, $y(t)$, and the drive, $x(t)$, systems.  
For the value of the control parameter
$b_2<0.97$, the similarity function become zero ($S_a(b_2)=0$) indicating that the
response system  anticipates the inverse state of the drive system confirming the existence of IAS.

It is to be noted that
one can observe exact inverse  anticipatory synchronization in the present study
even for the choice of the parameter $b_2$  for which only the less
stringent stability condition (\ref{piecewise_stab}) is satisfied in contrast to our earlier
studies~\cite{dvskml2005}, where one can observe only approximate direct anticipatory
synchronization under this condition
while exact synchronizations are observed for the values of the parameters
satisfying the general stability condition
for the coupled piecewise linear time-delay systems.  This is due to the fact that
in the present study, we have chosen the value of the delay time in the
drive system, $x(t)$, as $\tau=8$, for which the system has only two
positive Lyapunov exponents (as may be observed from the Fig.~\ref{fig1}).
This implies that the number of transversely unstable manifolds is less
and hence they are all stabilized even for the least values of the parameters
satisfying even the less stringent stability condition and so
one can observe exact inverse synchronization. This is in contrast to our earlier studies
on direct anticipatory synchronization~\cite{dvskml2005}, where we have chosen  $\tau=25$,
for which the drive system exhibits more than seven positive Lyapunov
exponents. Correspondingly there exists a large number of transversely unstable manifolds and 
hence the general stability condition is required to be satisfied in order to stabilize 
all the transversely unstable manifolds to obtain exact synchronization.

\begin{figure}
\centering
\includegraphics[width=1.0\columnwidth]{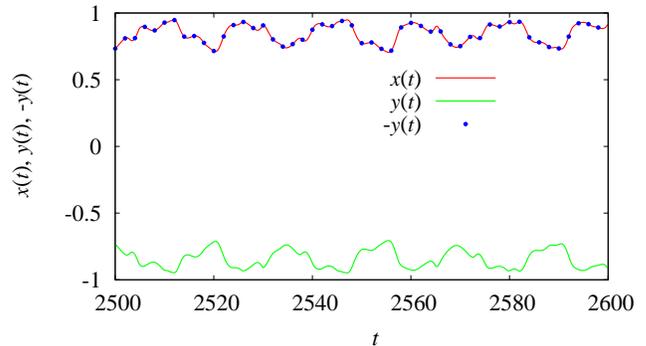}
\caption{\label{fig4} (Color online) The time trajectory of the variables
$x(t), y(t)$ and $-y(t)$ of the coupled piecewise linear time-delay systems indicating ICS
for the coupling delay $\tau_2=\tau_1=8.0$, while the other
parameter values are the same as in Fig.~\ref{fig2}.}
\end{figure}

As inverse anticipatory synchronization is a special case of generalized synchronization, 
the existence of IAS can also be further confirmed using the auxiliary system approach 
by augmenting the coupled piecewise linear time-delay systems ((\ref{eq.one}) and (\ref{eqoneb})) with an additional
auxiliary system for the variable $z(t)$ identical to the response system, satisfying the
equation
\begin{equation}
\dot{z}(t)=-az(t)+b_{2}f(z(t-\tau_{1}))-b_{3}f(x(t-\tau_{2})).
\label{aux}
\end{equation}

Now, the existence of IAS can be characterized by using the probability 
of synchronization~\cite{dvskml2005}, $\Phi(b_2)$,
calculated between the response, $y(t)$, and the auxiliary systems, $z(t)$, which
can be defined as the fraction of time during which $\left|y(t)-z(t)\right|<
\epsilon$ occurs, where $\epsilon$ is a small but arbitrary threshold.
The probability of synchronization, $\Phi(b_2)$, remains zero for the
value of the control parameter $b_2\ge 0.97$ as shown in Fig.~\ref{fig3}b, where
there is no synchronization between the response, $y(t)$, and the auxiliary system, $z(t)$.
On the other hand, for $b_2<0.97$, the probability of synchronization attains the value of unity
for the chosen threshold value for $\epsilon$,
clearly indicating the existence of complete synchronization between
the response, $y(t)$, and the auxiliary system, $z(t)$
(we have fixed the threshold value at $\epsilon=10^{-10}$ throughout the manuscript).
Correspondingly, there exits IAS between the coupled drive, $x(t)$, 
and the response, $y(t)$, systems with inhibitory coupling. 

\begin{figure}
\centering
\includegraphics[width=1.0\columnwidth]{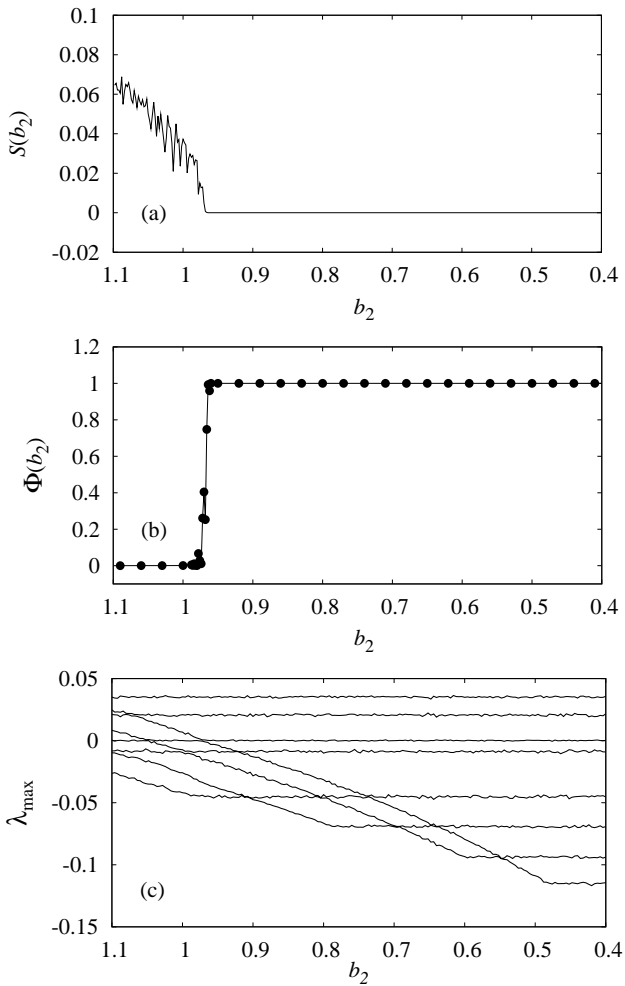}
\caption{\label{fig5} (a) Similarity function, $S_c(b_2)$, (b) Probability of
synchronization, $\Phi(b_2)$ and (c) First eight maximal Lyapunov exponents, $\lambda_{max}$,
of the coupled piecewise linear time-delay systems as a function of the
control parameter $b_2$, indicating the existence of ICS for $b_2<0.97$
and for the coupling delay $\tau_2=\tau_1=8.0$.}
\end{figure}

Further, the existence of IAS can also be characterized by the changes in the spectrum of
the Lyapunov exponents of the coupled systems, (\ref{eq.one})-(\ref{eqoneb}).
The first eight largest Lyapunov exponents of
the coupled piecewise linear time-delay system is shown in Fig.~\ref{fig3}c.
The two largest Lyapunov exponents of the drive system, $x(t)$, remain
positive, while the ones corresponding to the response system, $y(t)$, decrease
in their value as a function of the parameter $b_2$. The least positive
Lyapunov exponent of the response system becomes negative at $b_2=1.06$, while
the largest positive Lyapunov exponent of the response system becomes negative
at the value of $b_2=0.97$  confirming the onset of IAS at $b_2=0.97$.

\begin{figure}
\centering
\includegraphics[width=1.0\columnwidth]{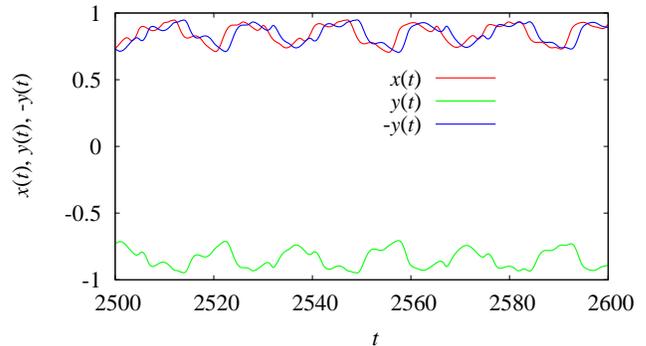}
\caption{\label{fig6} (Color online) The time trajectory of the variables
$x(t), y(t)$ and $-y(t)$ of the coupled piecewise linear time-delay systems indicating ILS
for the value of the coupling delay $\tau_2=10.0$, while the other
parameter values are the same as in Fig.~\ref{fig2}.}
\end{figure}

\subsection{Inverse Complete synchronization, ICS}

The synchronization manifold, $\Delta=x_{\tau_2-\tau_1}+y$,
becomes an ICS manifold for $\tau_1=\tau_2=8.0$. The time trajectory 
of the variables $x(t), y(t)$ and $-y(t)$  are plotted
in Fig.~\ref{fig4}, depicting the existence of ICS for the
same values of the other parameters as in Fig.~\ref{fig2}. 
It is clear from the Fig.~\ref{fig2} that the
state of the response system, $y(t)$, evolves in synchrony with the inverse state of 
the drive system $x(t)$. For the ease of clear visualization of this phenomenon, 
the inverse of the response $-y(t)$ is shown as filled circles in Fig.~\ref{fig2} 
and hence the inverse state of the response $-y(t)$  clearly evolves in synchrony 
with the state of the drive $x(t)$, there by 
illustrating the existence of ICS between the drive, $y(t)$, and the response, $x(t)$, systems. 

We have also
calculated the similarity function for the ICS defined as
\begin{eqnarray}
S_c^2(b_2)=\frac{\langle[y(t)+x(t)]^2\rangle}
{[\langle x^2(t)\rangle\langle y^2(t)\rangle]^{1/2}},
\label{cs:sim}
\end{eqnarray}
as a function of the parameter $b_2$ in 
Fig.~\ref{fig5}a. 
The similarity function, $S_c(b_2)$, oscillates with a finite value above zero
for the value of the control parameter $b_2 \ge 0.97$ as shown in Fig.~\ref{fig5}a,
where there does not exist any correlation between the interacting systems. However,
it is evident from Fig.~\ref{fig5}a that the similarity function  becomes
zero for the values of $b_2<0.97$ indicating the existence of
exact ICS in the coupled piecewise linear time-delay system even
for the value of the parameter satisfying the less stringent stability
condition as discussed in the previous section. We have also
characterized the existence of ICS using the probability of synchronization ($\Phi(b_2)$)
as shown in Fig.~\ref{fig5}b, which clearly shows that the probability
of synchronization becomes unity for $b_2<0.97$ depicting the existence of ICS
between the coupled drive, $x(t)$, and the response, $y(t)$, systems with inhibitory coupling.
However, for $b_2\ge 0.97$, the value of the probability of synchronization becomes
zero, $\Phi(b_2)=0$, indicating that there exist no correlation between the coupled
systems for this range of parameters.
The eight largest Lyapunov exponents of the coupled time-delay systems is
shown in Fig.~\ref{fig5}c. It shows that the two largest Lyapunov
exponents of the drive system remain positive while the largest 
positive Lyapunov exponent of the response system attains a negative value at
$b_2=0.97$, confirming the existence of ICS for $b_2<0.97$.

\begin{figure}
\centering
\includegraphics[width=1.0\columnwidth]{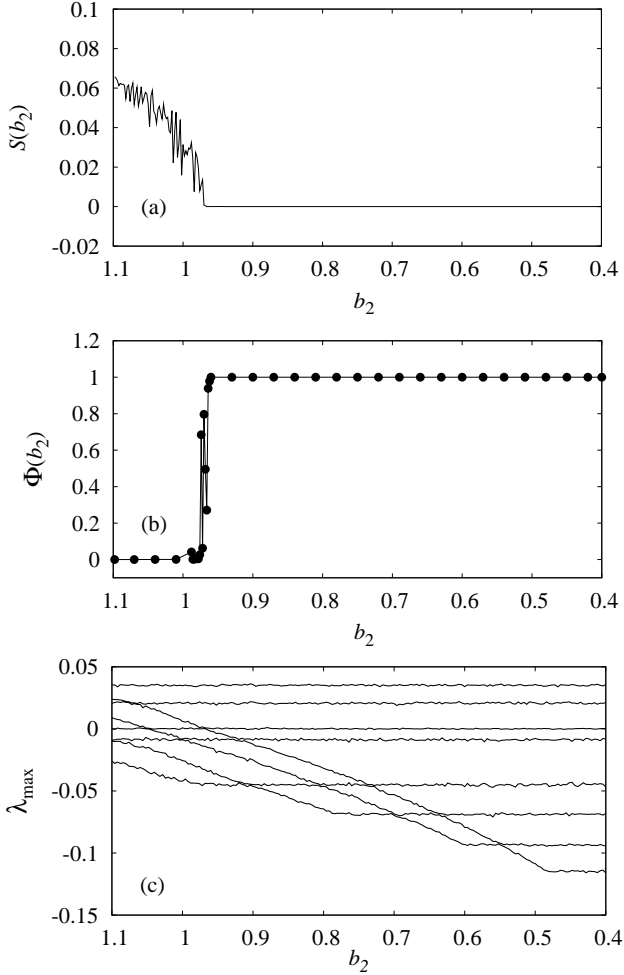}
\caption{\label{fig7} (a) Similarity function, $S_l(b_2)$, (b) Probability of
synchronization, $\Phi(b_2)$ and (c) First eight maximal Lyapunov exponents, $\lambda_{max}$,
of the coupled piecewise linear time-delay systems as a function of the
control parameter $b_2$, indicating the existence of ILS for $b_2<0.97$ and for the
coupling delay $\tau_2=10.0$.}
\end{figure}

\begin{figure}
\centering
\includegraphics[width=1.0\columnwidth]{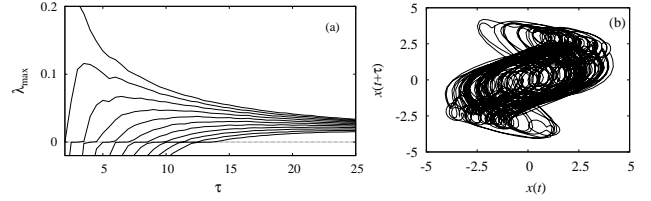}
\caption{\label{fig8} (a) The first eleven maximal Lyapunov exponents
$\lambda_{max}$ of the Ikeda time-delay system ((\ref{eq.one}a) and (\ref{eqonec})) for 
 $a=1.0, b=5$  $\tau\in(2,25)$ and (b) Hyperchaotic
attractor for the delay time $\tau_1=\tau=4.0$  with three positive Lyapunov
exponents for the above values of the other parameters.}
\end{figure}

\subsection{Inverse lag synchronization, ILS}

Now, we will demonstrate the existence of ILS in the
coupled piecewise linear time-delay system, (\ref{eq.one})-(\ref{eqoneb}), for the value of
coupling delay $\tau_2>\tau_1$ and for the same values of
the other parameters as in Sec.~\ref{ias}. The time series of the
variables $x(t), y(t)$ and $-y(t)$ of the coupled systems are shown in 
Fig.~\ref{fig6} for  $b_2=0.6$ and $\tau_2=10.0$, depicting the
existence of ILS in the coupled systems. It is evident from the figure that the
state of the response system $y(t)$ lags the inverse state of the drive system $x(t)$.
In order to made this clear visually, the inverse state of the response $-y(t)$ is 
plotted in Fig.~\ref{fig6} and hence
the inverse state of the response $-y(t)$ lags the state of the drive $x(t)$, there by 
illustrating the existence of ILS between the drive and the response systems. 

The similarity function for ILS defined as
\begin{eqnarray}
S_l^2(b_2)=\frac{\langle[y(t+\tau)+x(t)]^2\rangle}
{[\langle x^2(t)\rangle\langle y^2(t)\rangle]^{1/2}},
\label{lag:sim}
\end{eqnarray}
is plotted in Fig.~\ref{fig7}a as a function of  $b_2$. 
The similarity function oscillates with a finite amplitude for $b_2 \ge 0.97$
as in the other cases, as there is no synchronous evolution among the coupled systems.
However, the similarity function, $S_l(b_2)$, reaches zero for $b_2<0.97$, confirming
the existence of ILS in the coupled time-delay systems. Similarly, the value of the probability
of synchronization, $\Phi(b_2)$, calculated between the response, $y(t)$, and the auxiliary,
$z(t)$, systems attains  unity for $b_2<0.97$, confirming the existence
of complete synchronization between them as shown in Fig.~\ref{fig7}b.  Correspondingly, 
there exists ILS between the
drive, $x(t)$, and the response, $y(t)$, systems. The two largest Lyapunov  exponents of 
the drive system remain positive as shown in Fig.~\ref{fig7}c, while the largest 
positive Lyapunov exponent of the response system becomes negative at $b_2=0.97$, confirming
the existence of ILS between the drive and the response systems.

\begin{figure}
\centering
\includegraphics[width=1.0\columnwidth]{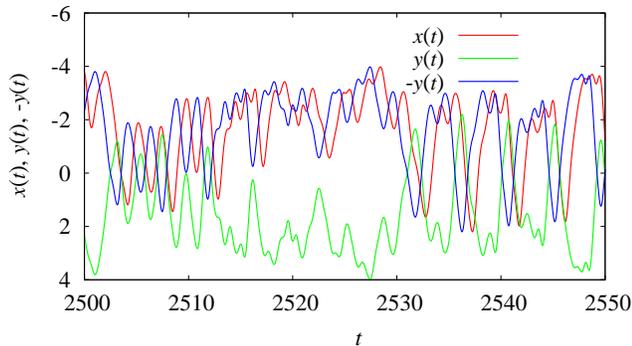}
\caption{\label{fig9} (Color online) The time trajectory  of the variables
$x(t), y(t)$ and $-y(t)$ for $a=1.0, b_1=5.0,b_2=2.0, b_3=3.0, \tau_1=4.0$  and $\tau_2=3.0$
of the coupled Ikeda time-delay systems indicating IAS.}
\end{figure}

\section{Inverse synchronizations in the coupled Ikeda time-delay systems ((\ref{eq.one}) and (\ref{eqonec}))}

In this section, as an illustration for the case where the coefficient of the
$\Delta$ term  in the error equation (\ref{eq.difsys}) is constant, while that of
the $\Delta_\tau$ term is time-dependent,
we will demonstrate the existence of inverse anticipatory, inverse complete and 
inverse lag synchronizations as a function of the coupling delay in the
coupled Ikeda time-delay systems (\ref{eq.one}) and (\ref{eqonec})
with inhibitory coupling.

\subsection{\label{i_ias}Inverse anticipatory synchronization, IAS}

We have fixed the values of the parameters of the coupled Ikeda time-delay systems ((\ref{eq.one}) and (\ref{eqonec}))
as $a=1.0, b_1=5, \tau_1=4.0$ and $\tau_2=3.0$, while the other two parameters
$b_2$ and $b_3$ are fixed according to the parametric condition $b_1=b_2+b_3$.
To appreciate the chaotic and hyperchaotic nature of the uncoupled system, we 
present in Fig.~\ref{fig8}a the first eleven largest Lyapunov exponents for the 
above values of the parameters in the range of delay time $\tau\in(2,25)$ where
several of them take positive values. As an illustration, the hyperchaotic 
attractor with three positive Lyapunov exponents of the uncoupled system for 
$\tau_1=\tau=4.0$ is depicted in Fig.~\ref{fig8}b.

\begin{figure}
\centering
\includegraphics[width=1.0\columnwidth]{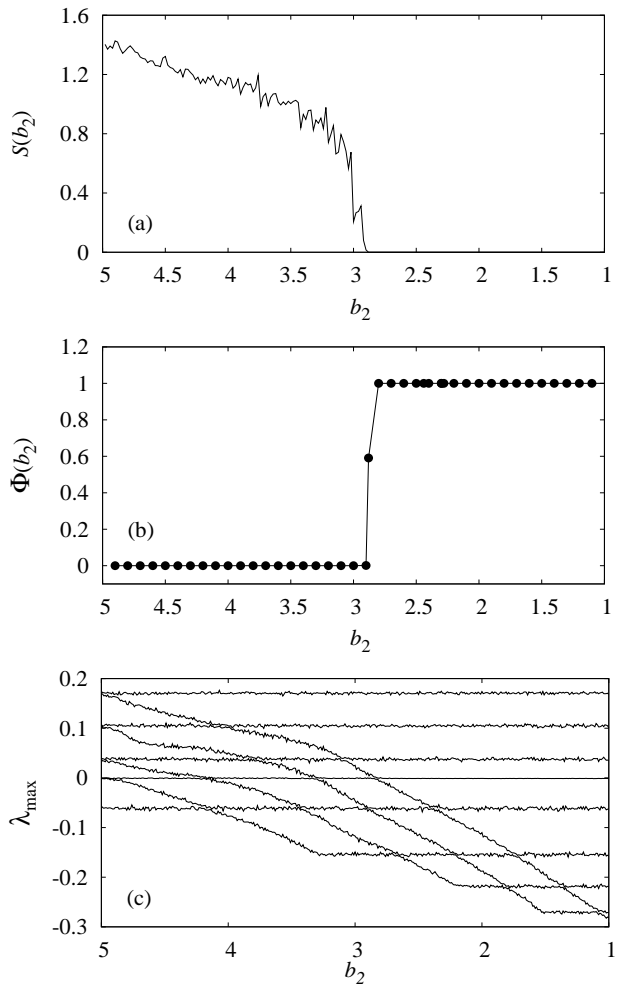}
\caption{\label{fig10} (a) Similarity function, $S_l(b_2)$, (b) Probability of
synchronization, $\Phi(b_2)$ and (c) First nine maximal Lyapunov exponents, $\lambda_{max}$,
of the coupled Ikeda time-delay systems as a function of the
control parameter $b_2$, indicating the existence of IAS for $b_2<2.88$ and for the
coupling delay $\tau_2=3.0$.}
\end{figure}

Now, we will demonstrate the existence of IAS for the value of the coupling delay
$\tau_2$ less than that of the feedback delay $\tau_1$  and for the values of the other 
parameters satisfying the stability condition (\ref{ikeda_stab}). It is evident
from Fig.~\ref{fig8}b that the maximum value of $x(t)$ does not exceed $x_{max}=5$.
Correspondingly $x(t-\tau_2)_{max}=5$. As a consequence the stability condition 
(\ref{ikeda_stab}) can be written as
\begin{align}
a>b_2\cos(5)=0.284 b_2.
\label{ikeda_asy}
\end{align}
Then, one can obtain an asymptotically stable synchronized
state for the values of parameters satisfying the above stability condition. The 
time trajectories of the variables
$x(t), y(t)$ and $-y(t)$ depicting the existence of IAS  are plotted in Fig.~\ref{fig9}
for the value of the parameter $b_2=2.0$ satisfying the stability condition (\ref{ikeda_asy}).
The minimum of the similarity function, $S_a(b_2)$, defined by Eq.~(\ref{anti:sim}) becomes zero
for $b_2<2.88$ as shown in Fig.~\ref{fig10}a, indicating the existence of IAS
in the coupled Ikeda time-delay system.

\begin{figure}
\centering
\includegraphics[width=1.0\columnwidth]{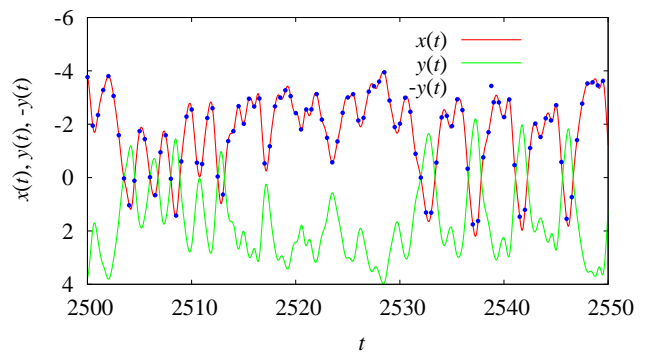}
\caption{\label{fig11} (Color online) The time trajectory of the variables
$x(t), y(t)$ and $-y(t)$ for $a=1.0, b_1=5.0,b_2=2.0, b_3=3.0, \tau_1=4.0$  and $\tau_2=4.0$
of the coupled Ikeda time-delay systems indicating ICS.}
\end{figure}

The existence of IAS is further characterized by the probability of synchronization, 
$\Phi(b_2)$, for complete synchronization between the response,
$y(t)$, and the auxiliary, $z(t)$, systems by augmenting the coupled Ikeda time-delay 
systems ((\ref{eq.one}) and (\ref{eqonec})) with an additional
auxiliary system for the variable $z(t)$ identical to the response system as in Eq.~(\ref{aux}).
Figure.~\ref{fig10}b
indicates that the value of the probability of synchronization becomes $\Phi(b_2)=1.0$
for $b_2<2.88$, confirming the existence of IAS between the coupled drive, $x(t)$, and the response, $y(t)$,
systems. The existence of IAS for $b_2<2.88$ is further confirmed from the changes in the
spectrum of the largest Lyapunov exponents of the coupled Ikeda time-delay systems ((\ref{eq.one}) and (\ref{eqonec})).
The nine largest Lyapunov exponents, $\lambda_{max}$, is shown in Fig.~\ref{fig10}c
as a function of $b_2$. While the largest three positive Lyapunov exponents of the 
drive system remain positive, the largest positive exponents of the response system gradually become
negative as a function of $b_2$ and finally they 
become negative at $b_2=2.88$, confirming the existence of IAS for $b_2<2.88$ satisfying the
stability condition (\ref{ikeda_asy}).

\begin{figure}
\centering
\includegraphics[width=1.0\columnwidth]{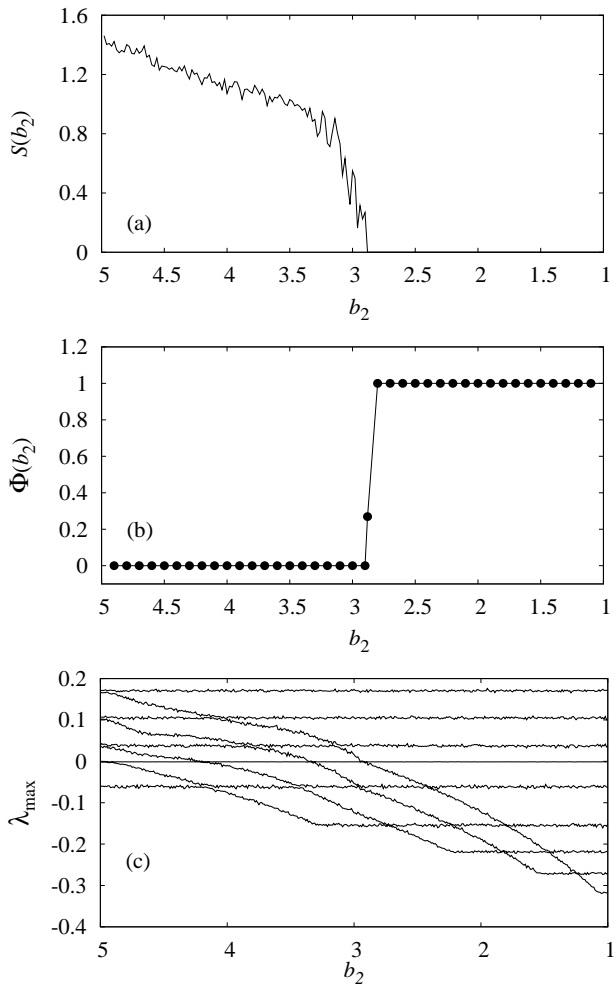}
\caption{\label{fig12} (a) Similarity function, $S_l(b_2)$, (b) Probability of
synchronization, $\Phi(b_2)$ and (c) First nine maximal Lyapunov exponents, $\lambda_{max}$,
of the coupled Ikeda time-delay systems as a function of the
control parameter $b_2$, indicating the existence of ICS for $b_2<2.88$ and for the
coupling delay $\tau_2=\tau_1=4.0$.}
\end{figure}

\subsection{Inverse complete synchronization, ICS}

The time series of the variables $x(t), y(t)$ and $-y(t)$ indicating the 
existence of ICS, for the value of the coupling
delay $\tau_2=\tau_1=4.0$, and for the same values of the other parameters satisfying the
stability condition (\ref{ikeda_asy}) as in the previous section,
are shown in Fig.~\ref{fig11}.
The similarity function, $S_c(b_2)$, given by (\ref{cs:sim}), for ICS shown in Fig.~\ref{fig12}a indicates that
the minimum of $S_c(b_2)=0$ for $b_2<2.88$ depicting the existence of ICS  in the corresponding
range of $b_2$. It is also evident from the  probability  of synchronization (Fig.~\ref{fig12}b)
that $\Phi(b_2)=1$ for $b_2<2.88$ confirming the existence of ICS between the
drive and the response systems.  This transition from a desynchronized state to ICS for the
values of parameters satisfying the stability condition (\ref{ikeda_asy}) is also
confirmed from the changes in the spectrum of the Lyapunov exponents of the coupled Ikeda systems as
shown in Fig.~\ref{fig12}c. The largest positive Lyapunov exponents of the
response system become negative at $b_2=2.88$ confirming the existence of ICS for $b_2<2.88$,
while the largest three positive Lyapunov exponents of the drive system remain positive. 

\begin{figure}
\centering
\includegraphics[width=1.0\columnwidth]{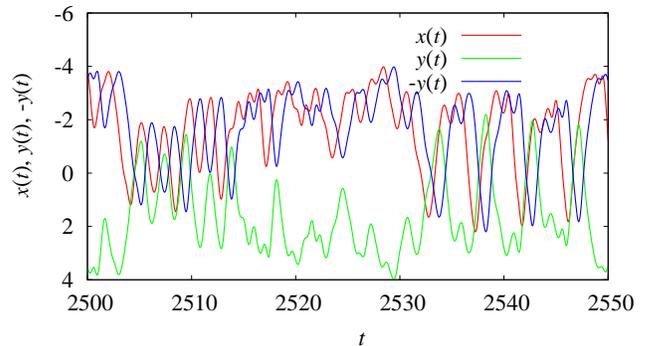}
\caption{\label{fig13} (Color online) The time trajectory of the variables
$x(t), y(t)$ and $-y(t)$ for $a=1.0, b_1=5.0,b_2=2.0, b_3=3.0, \tau_1=4.0$  and $\tau_2=5.0$
of the coupled Ikeda time-delay systems indicating ILS.}
\end{figure}

\subsection{Inverse lag synchronization, ILS}

\begin{figure}
\centering
\includegraphics[width=1.0\columnwidth]{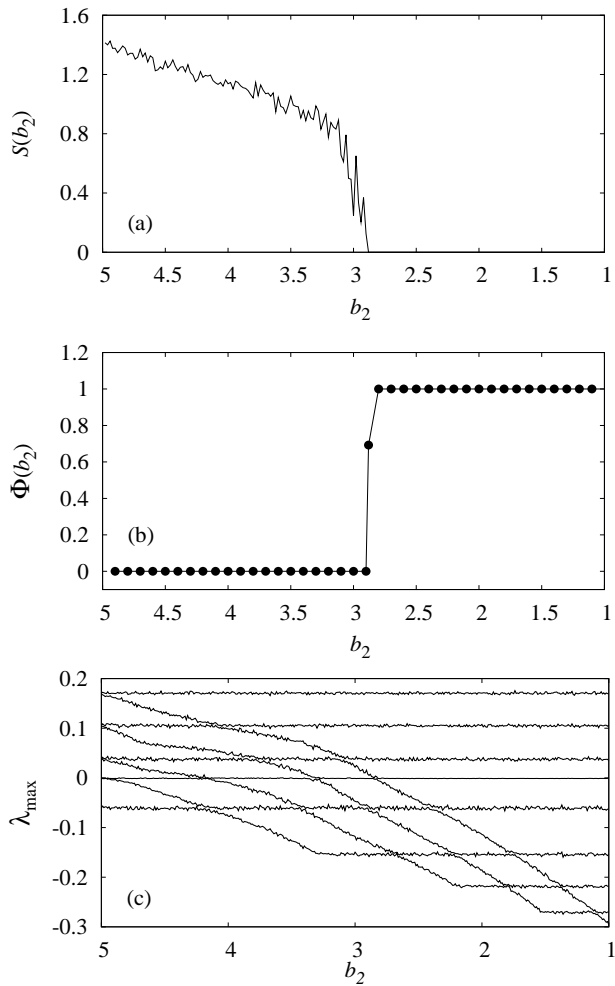}
\caption{\label{fig14} (a) Similarity function, $S_l(b_2)$, (b) Probability of
synchronization, $\Phi(b_2)$ and (c) First nine maximal Lyapunov exponents, $\lambda_{max}$,
of the coupled Ikeda time-delay systems as a function of the
control parameter $b_2$, indicating the existence of ILS for $b_2<2.88$ and for the
coupling delay $\tau_2=5.0$.}
\end{figure}

For  the coupling delay $\tau_2=5.0$ greater than the feedback delay
$\tau_1=4.0$, the synchronization manifold $\Delta=0$ corresponds to the ILS manifold.
The time trajectories of the variables $x(t), y(t)$ and $-y(t)$ of the coupled Ikeda
time-delay systems (Fig.~\ref{fig13}) clearly depict the ILS for the above choice of
the coupling delay (the other parameters are fixed as in Sec.~\ref{i_ias} satisfying
the stability condition (\ref{ikeda_asy})). The minimum
of the similarity function for ILS turns out to be $S_l(b_2)=0$ as shown in
Fig.~\ref{fig14}a for 
$b_2<2.88$ indicating the existence of ILS for $b_2<2.88$. Similarly, the value of the
probability of synchronization (Fig.~\ref{fig14}b) becomes unity in the corresponding range of $b_2$
confirming the existence of ILS between the coupled Ikeda time-delay systems. The existence of ILS
for $b_2<2.88$ is also further confirmed by the changes in the spectrum of the Lyapunov exponents of the
coupled Ikeda time-delay systems as shown in Fig.~\ref{fig14}c. The  largest three 
positive Lyapunov exponents of the drive system remain unchanged, while that of the response system decrease
in their values as a function of $b_2$ and they
become negative at $b_2=2.88$  confirming the emergence of ILS in coupled Ikeda time-delay
systems. 

\section{Summary and conclusion}
In this paper, we have shown the transition from inverse anticipatory
to inverse lag via inverse complete synchronization as a function of the
coupling delay $\tau_2$ for fixed value of the feedback delay $\tau_1$
in unidirectionally coupled time-delay systems with inhibitory coupling.
We have also arrived a suitable stability  condition for the
asymptotic stability of the synchronized states using the Krasovskii-Lyapunov
functional theory. We have demonstrated that the same general stability
condition resulting from Krasovskii-Lyapunov functional approach
can be valid for two different cases, where (i) both the coefficients of the
$\Delta$ and $\Delta_\tau$ terms of the error equation corresponding
to synchronization manifold are constants and (ii) the coefficient
of the $\Delta_\tau$ term is time dependent while that of the other is
time independent using suitable examples. 
The existence of different types of inverse synchronizations are corroborated
using similarity function, probability of synchronization and from the
changes in the spectrum of the largest Lyapunov exponents of the coupled
time-delay systems. We have also designed suitable couplings for the case where 
both coefficients of the $\Delta$ and $\Delta_\tau$ terms are time dependent
to show the validity of the same general stability condition (\ref{eq.asystab}) 
resulting from the Krasovskii-Lyapunov
functional theory, the results of that will be published in a forthcoming paper.

\begin{acknowledgments}
The work of DVS has been supported by Alexander von Humboldt Foundation.
JK has been supported by his Humboldt-CSIR research award and NoE BIOSIM (EU)
Contract No. LSHB-CT-2004-005137.
ML acknowledges the support from  a Department of Science and Technology (DST),
Government of India sponsored IRHPA research project and DST Raja Ramanna Program.
\end{acknowledgments}


\end{document}